\documentstyle[12pt,aps,multicol,pre,psfig]{revtex}
\begin{document}
\title{Exploring the Collective  Behavior on the One-dimensional Quartic Map}
\author{Leonardo~G.~Brunnet\cite{emaileon}, Jason A.~C.~Gallas\cite{bylinejg}}
\address{Instituto de F{\'\i}sica, Universidade Federal do Rio Grande 
      do Sul, 91501-970 Porto Alegre, Brazil}
\date{\today}
\maketitle
\begin{abstract}
In this work we study numerically a lattice composed
of  two parameter single quartic maps with local diffusive
coupling. We find large regions over the parameter space where the
single quartic map is periodic and the coupled system is not
synchronized. We explore further one point on this region to show that
it presents a real non-trivial collective behavior. 


\end{abstract}
\section{Introduction}

Since the last decade different systems composed of coupled map
lattices (CML) have been studied as simple toy models used to unveil basic
properties of spatially extended nonlinear dynamical systems.
Conceived as a far from equilibrium system with many degrees of
freedom such systems can also be used to construct a statistical
mechanics for irreversible process.

In the beginning of the decade, studying systems of locally coupled
logistic maps on spatial dimensions greater than one, Chat{\'e} and
Manneville \cite{chate92} found a temporal dynamics on spatial
averages over the system emerging out of the local chaos, the so
called {\it non-trivial collective behavior} (NTCB). They found that
such behavior happens when the single logistic map parameter is set
to any value beyond the chaotic onset, that is for
$\lambda~>~\lambda_c$. When $\lambda~<~\lambda_c$ the whole system
synchronizes to the local periodic behavior. So, the idea here is that
the basic ingredients to find NTCB are local chaos competing with
local sensibility to initial conditions (SIC).

Most works found in the literature \cite{kaneko} use the logistic map
to describe the local dynamics. A characteristic feature of the
logistic map is the presence a single attractor on the interval of
convergence of its variable. 
On a CML
composed of such units the effect of the local coupling is to average
the variable (field) over the neighborhood producing a new value
inside the basin of attraction of the very same attractor.
The motivation of this work is to investigate a CML possessing two
non-infinite attractors. In order to do that we use for the local
dynamics the one-dimensional quartic map introduced by one of us in
1993 \cite{jasonPA94}. 

%
%

Starting from random initial conditions we ran a system composed of
$N$ units whose  dynamics at the site $i$ follows the quartic map,
$${y_i}^{t+1}=((x_i^{t+1})^2 - a)^2 - b, $$
 where $x_i^t$ is a real variable representing some quantity of
interest measured at time $t$; $a$ and $b$ are real parameters. After
each generation $t$ the value of $y_t$ is averaged over the first
neighbors, $$x_i^t = (1-\epsilon) {y_i}^t + {\epsilon \over 2D}
\sum_{j=1}^{n}~{y_j^{t}}$$ where $D$ is the spatial dimension of the
system and $n$ is the number of nearest neighbors. The geometry used for
the simulation presented here is of a square lattice of sizes up to $N
= 512\times 512$ units.

We first explored the system scanning the space of parameters $a$ and
$b$ on the interval $\{-1,2\}$ dividing it with a resolution of $512
\times 512$. For each couple of parameters a small lattice of $64
\times 64 $ sites was iterated up to one thousand times departing from
initial conditions randomly chosen from the basin of attraction of the
uncoupled quartic map. From such experiment we could separate the set
of parameters whose lattice rapidly synchronize from those which
present either a long transient or a non-trivial collective-behavior
(NTCB). In figure 1 we show these results in connection with
the Lyapunov exponents of the local map. In this figure the black
regions show sets of parameters where the coupled system does not
synchronize up to one thousand iterations {\it and} the local
uncoupled map is periodic, the white color represents regions where
the system does not synchronize {\it and} the local uncoupled map is
chaotic, the dark gray shows the regions where the coupled system
synchronizes to the local uncoupled map and the pale gray color shows
the orbits which are attracted to infinite.

The striking result is shown in the black region of figure 1, there we
may have NTCB without chaos on the local uncoupled oscillator. In fact
the absence of synchronization of the variable over the space may also
be just a long transient. To clear out this we have particularly
explored the point $a=0.35, b=1.35$ on the black region. The local
uncoupled map has two fixed points, $x_1=-1.227591$, $x_2=-0.011399$
and negative Lyapunov exponents $L_1=L_2=-1.200$. The coupled system
presents a period two NTCB, after a small transient the instantaneous
average over the variable jumps alternatively from the interval
$\{.125144,.815711\}$ to the interval $\{-1.349951,-1.245386\}$. See
figure 2 \ref{fig2}. We have
performed $2.5 \times 10^5$ iterations with large lattices of $512
\times 512$ sites and up to $10^6$ interactions with smaller lattices
of $128 \times 128$ sites without observing synchronization on the
system. In figure \ref{fig2} we show the images of a lattice of $512
\times 512$ sites in two consecutive steps after $2.5 \times 10^5$
iterations.


In order to confirm the NTCB for this point of the parameter space we
have observed the dependence of the time fluctuation of the spatial
average of the field variable with the lattice size during the same
time interval.  For large systems this average is expected to become
more and more defined \cite{chate92}. In fact for increasing lattice
sizes the fluctuation decreases with the $\sqrt{N}$ \ref{fig3} and the
NTCB is better and better defined. Another test to the NTCB is to
verify if it robust to changes on the parameters, so we ran the system
on a small region of radius $0.01$ around this point on the parameter
space. We have found the same period two NTCB with small changes on
the spatial averages showing that it is robust to small changes on
parameters.



From the first work of Chat\'e and Manneville \cite{chate92} it was
already clear that the NTCB existed for the logistic map on the so
called periodicity windows appearing when $\lambda~>~\lambda_c$. A similar
result has also been remarked by one of us in a collaboration with
Chat\'e and Manneville~\cite{leo94} when exploring the robusticity of
the NTCB on the coupled R\"ossler system. In both cases, nevertheless,
the NTCB for periodic parameters emerges on a small window on the
parameter space surrounded by chaos. Here it emerges on the boarding
region between periodic and chaotic behavior.

From these results we can state that sensibility to initial conditions
(SIC) on the local attractor is not a necessary condition to NTCB and
we may find NTCB even when the system is periodic over large regions
of the parameter space. Nevertheless the existence of spatial
dispersion clearly indicates that the diffusive coupling, which forces
the system to synchronize, is counterbalanced by some kind of
SIC. What means that knowledge of the local attractor does not implie
knowledge of the global one which emerges from the coupled system.  We
may try to generalize summing up the simulations presented here and
the previous results. That is, for the NTCB to exist on a periodic
local base we need either a basin of attraction more complex with two
relevant parameters (and two attractors) or a small window of
periodicity surrounded by chaos.
\acknowledgments
We thank H.~Chat\'e and A. Pikovsky for important discussions and
suggestions. We acknowledge the Centro de Supercomputa{\c c}{\~a}o of
our university for the computing time on the CRAY YP-MP2 and the
supporting Brazilian agencies CNPq and FAPERGS.

\newpage
{\Large Figure 1}
\begin{figure}[h] \vbox{
\centerline{\psfig{figure=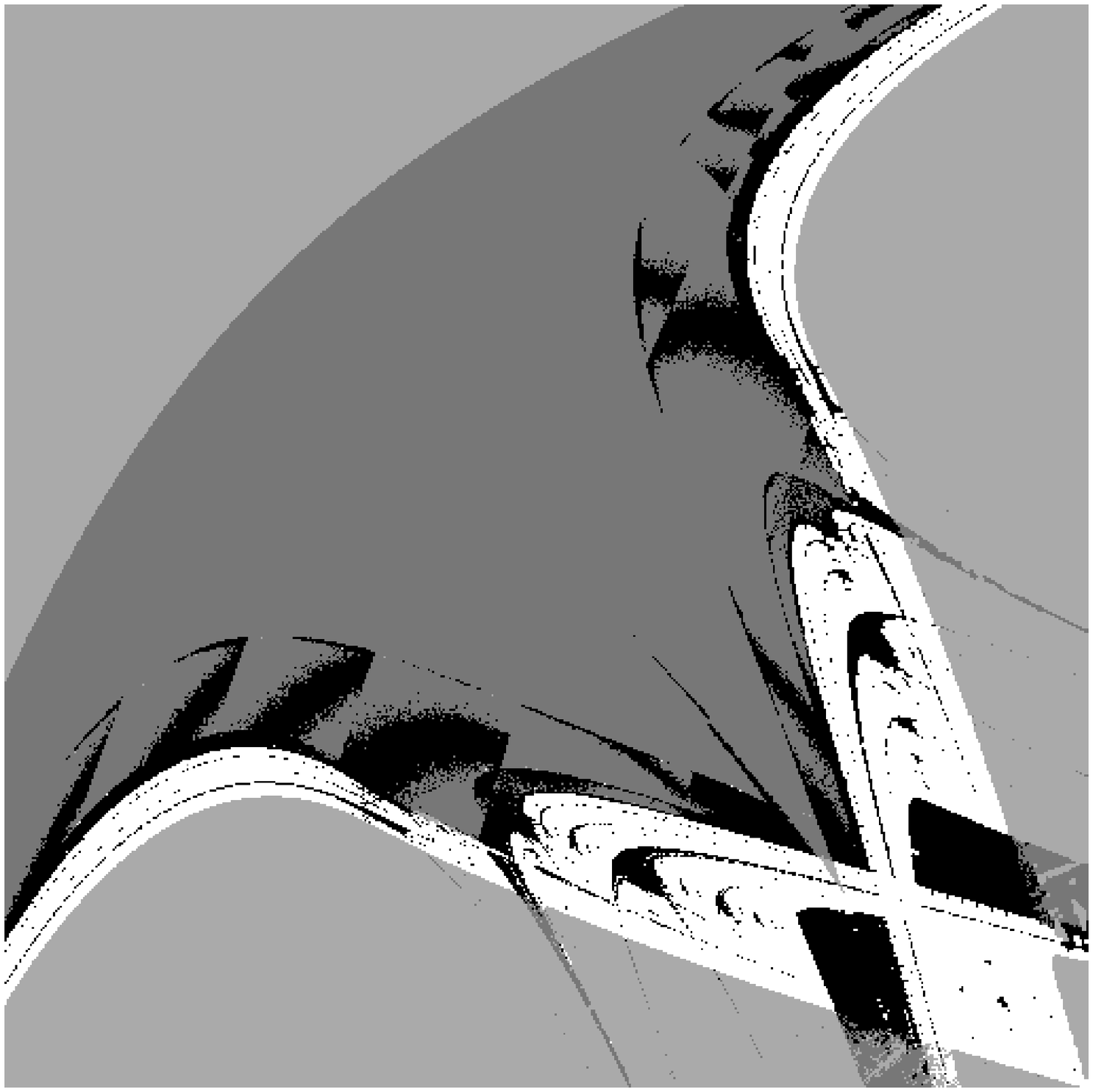,width=17cm,angle=0}}
\label{fig1}
}\end{figure}
\newpage
{\Large Figure 2}
\begin{figure}[] \vbox{
\centerline{\psfig{figure=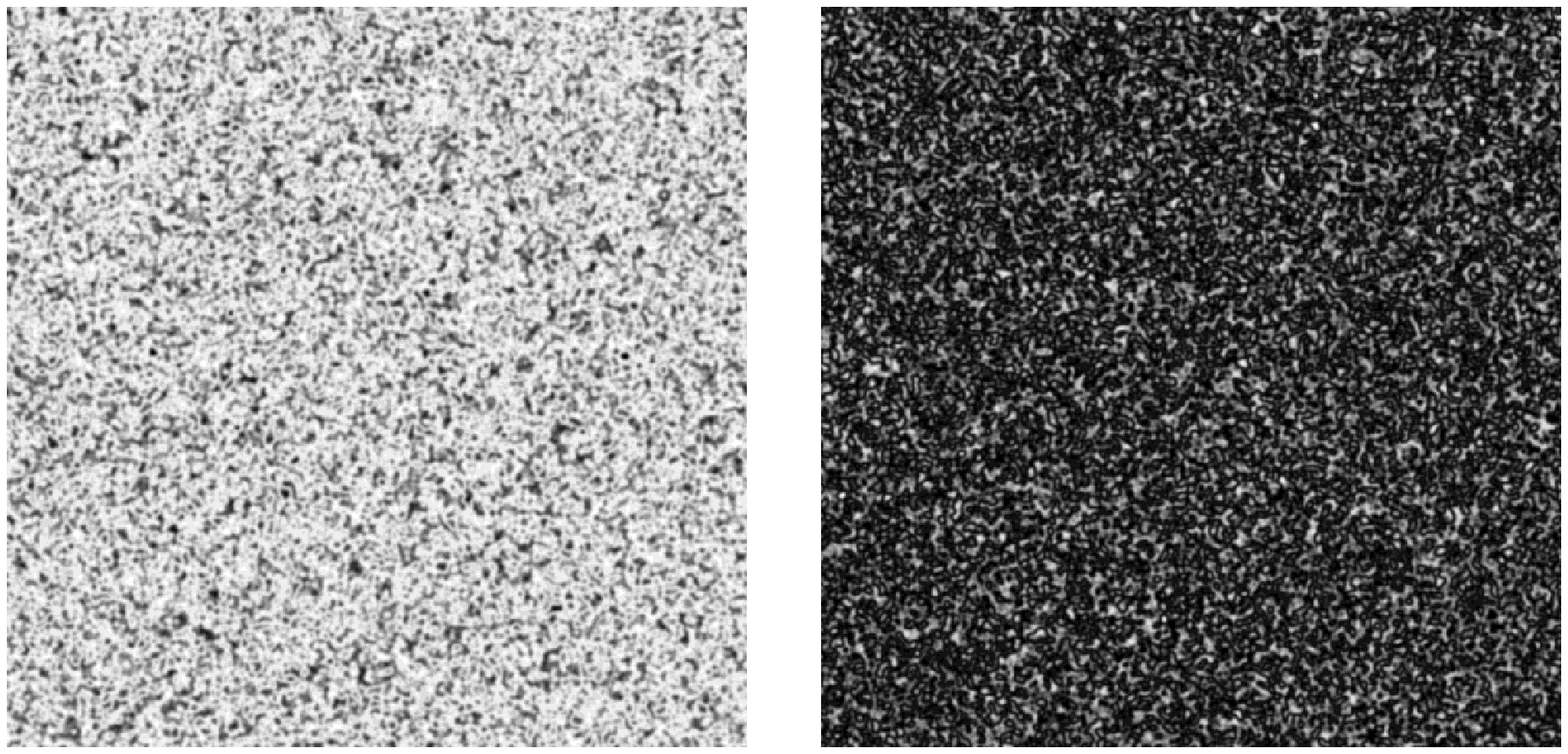,width=11cm,angle=90}}
\label{fig2}
}\end{figure}
\newpage
{\Large Figure 3}
\begin{figure}[] \vbox{
\centerline{\psfig{figure=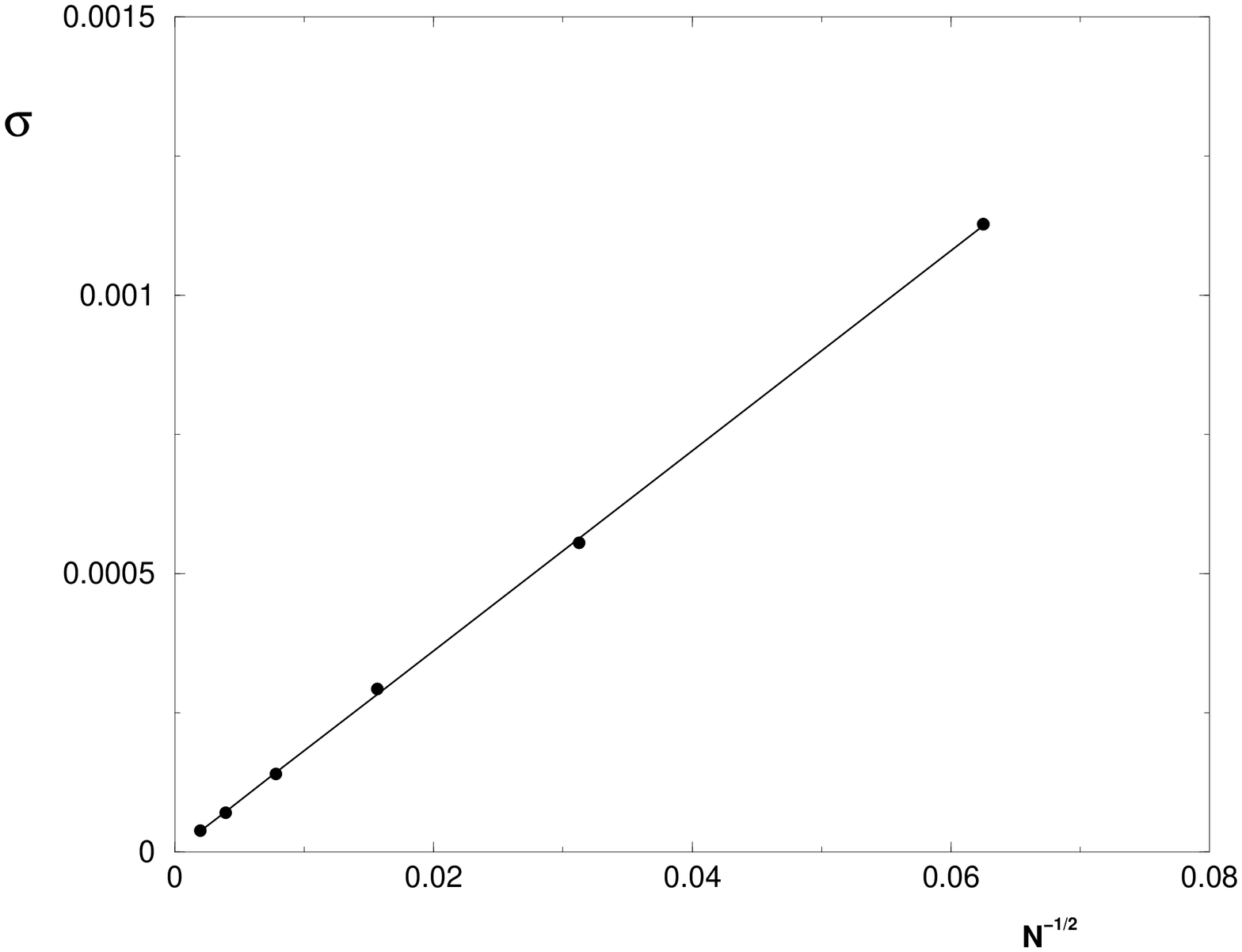,width=17cm,angle=270}}
\label{fig3}
}
\end{figure}

\newpage
\noindent{\Large {\bf Figure Captions}}\\
\underline{Figure 1}\\
{Image of parameter space. Black: region where the single
quartic map is periodic {\it and} the coupled system does not
synchronize. White: the single map is chaotic and the coupled system
does not synchronize. Dark gray: coupled system synchronizes to the
local uncoupled map. Pale gray: orbits attracted to infinite. The
scale in both axes is in the interval $\{-1.0,2.5\}$.}\\
\underline{Figure 2}\\
{Image of the spatial variable ($x$)  showing a period two NTCB for
parameters a=0.35 and b=1.35. The gray level scale on left is  in the interval $\{0.125144,0.815711\}$
and on the right in the interval $\{-1.349951,-1.245386\}$.}\\
\underline{Figure 3}\\
{Time fluctuation of the lattice spatial average, mean deviation for different  lattice sizes.}


\begin{references}
\bibitem[+]{bylinejg}  Electronic mail: jgallas @ if.ufrgs.br
\bibitem[*]{emaileon}  Electronic mail: leon @ if.ufrgs.br
\end{references}
\end{document}